\documentclass{ptephy_v2}%%%%where ptephy_v1 is the template name
\usepackage{hyperref}
\usepackage{stackengine}
\begin{document}

\title{{\large\tt\bf Letter}\\ \vskip+0.15cm
Cross section Measurements for $^{12}$C$(K^-, K^+\Xi^-)$ and $^{12}$C$(K^-, K^+\Lambda\Lambda)$ Reactions at 1.8 GeV$/c$}
%%%% To generate auto affiliation numbers please use \author{}\affil{} command

\author{Woo Seung Jung}
\author[2,6]{Yudai Ichikawa}
\author[1]{Byung Min Kang}
\author[1]{Jung Keun Ahn$^\ast$}
\affil{\normalsize Department of Physics, Korea University, Seoul 02841, Republic of Korea 
\email{ahnjk@korea.ac.kr}}
\author[1]{Sung Wook Choi} 
\author[6]{Manami Fujita}
\affil{\normalsize Department of Physics, Tohoku University, Sendai 980-8578, Japan}
\author[3,6]{Takeshi Harada}
\affil{\normalsize Department of Physics, Kyoto University, Kyoto 606-8502, Japan}
\author[6]{Shoichi Hasegawa}
\author[2]{Shuhei Hayakawa}

\author{Sang Hoon Hwang} 
\affil{\normalsize Korea Research Institute of Standards and Science, Daejeon 34113, Korea}

\author{Kenneth Hicks} 
\affil{\normalsize Department of Physics \& Astronomy, Ohio University, Athens, Ohio 45701, USA}

\author[3,6]{Ken'ichi Imai}

\author[2]{Yuji Ishikawa} 
\affil{\normalsize Advanced Science Research Center, Japan Atomic Energy Agency, Tokai 319-1195, Japan}

\author[2]{Shunsuke Kajikawa} 
\author[2]{Kento Kamada}

\author{Shin Hyung Kim} 
\affil{\normalsize Department of Physics, Kyungpook National University, Daegu 41566, Republic of Korea}

\author[2]{Tomomasa Kitaoka} 
\author{Jaeyong Lee}
\affil{\normalsize Department of Physics and Astronomy, Seoul National University, Seoul 08826, Republic of Korea}
\author[1]{Jong Won Lee}

\author[2,9]{Koji Miwa}
\author[2]{Taito Morino} 
\author[2]{Fumiya Oura}
\author[6]{Hiroyuki Sako}
\author[2]{Tamao Sakao} 
\author[2]{Masayoshi Saito}
\author[6]{Susumu Sato}
\author[2,6]{Hirokazu Tamura}
\author[6]{Kiyoshi Tanida}

\author{Toshiyuki Takahashi} 
\affil{\normalsize Institute of Particle Nuclear Study, High Energy Accelerator Research Organization (KEK), Tsukuba 305-0801, Japan}

\author[2,9]{Mifuyu Ukai}
\author[2]{Shunsuke Wada} 
\author[6]{Takeshi O. Yamamoto}
\author[1]{Seongbae Yang}

%%% To include the collaborator name... Please use the command "\collaborator"
% \collaborator{J-PARC E42 Collaboration}

\begin{abstract}%
We present a measurement of the production of $\Xi^-$ and 
$\Lambda\Lambda$ in the $^{12}$C$(K^-, K^+)$
reaction at an incident beam momentum of 1.8 GeV$/c$, 
based on high-statistics data from J-PARC E42. 
The cross section for the $^{12}$C$(K^-, K^+\Xi^-)$ reaction, compared to the inclusive  
$^{12}$C$(K^-, K^+)$ reaction cross section, indicates that the $\Xi^-$ emission probability 
peaks at 70\% in the energy region of $E_\Xi=100$ to 150 MeV 
above the $\Xi^-$ emission threshold.
A classical approach using eikonal approximation shows that the total cross sections 
for $\Xi^-$ inelastic scattering ranges between 42 mb and 23 mb in the $\Xi^-$ momentum 
range from 0.4 to 0.6 GeV$/c$. 
Furthermore, based on the relative cross section for the $^{12}$C$(K^-, K^+\Lambda\Lambda)$ reaction, the total cross section for $\Xi^-p\to\Lambda\Lambda$ 
is estimated in the same approach to vary between 2.2 mb and 1.0 mb 
in the momentum range of 0.40 to 0.65 GeV$/c$. 
Specifically, a cross section of 1.0 mb in
the momentum range of 0.5 to 0.6 GeV$/c$
imposes a constraint on the upper bound of the decay width 
of the $\Xi^-$ particle in infinite nuclear matter, 
revealing $\Gamma_\Xi< \sim 0.6$ MeV. 
\end{abstract}
\subjectindex{D33}
\maketitle
%\linenumbers
\vskip-0.4cm
{\it 1. Introduction} 
The study of hyperon-nucleon interactions, particularly in the $S=-2$ system, 
has attracted significant attention due to its potential to provide 
insights into flavor SU(3) symmetry breaking in baryon-baryon interactions. 
However, data on the $S=-2$ system remain quite limited. 
Recently, the ALICE collaboration investigated the interaction between $\Xi^-p$ 
using femtoscopy measurements, which suggested an attractive interaction 
for the $\Xi N$ system \cite{alice}. Additionally, research on twin $\Lambda$ hypernuclei,
discovered in a nuclear emulsion experiment at J-PARC, 
indicated that the $\Xi N - \Lambda \Lambda$ coupling is weak \cite{e07}. 
This finding is consistent with results from a recent Lattice QCD calculation \cite{lattice}. 

The existence of $\Xi$ hypernuclei has been investigated using missing mass spectroscopy 
in the $^{12}$C$(K^{-},K^{+})$ reaction in the KEK E224\cite{e224}, BNL E885\cite{e885},
and J-PARC E05 experiments \cite{e05}.
The BNL experiment suggested that the depth of the $\Xi N$ potential is
approximately $V_{0}^{\Xi} \sim -14$ MeV, assuming a Woods-Saxon type potential. 
However, a clear peak structure was not observed due to limitations in energy resolution.
More recently, the J-PARC E05 reported a significant spectral enhancement near the
$\Xi^-$ production threshold. Furthermore, the J-PARC E70 experiment is currently 
collecting data to identify a distinct peak structure, utilizing a missing-mass
resolution of 2 MeV (FWHM) \cite{e70}.

Furthermore, a theoretical study that assumed 
a $\Xi N \rightarrow \Lambda \Lambda$ conversion width 
as $\Gamma/2 \approx 4 \ \text{MeV}$ 
concluded that the data from BNL E885 favors a potential strength closer to $V_0^\Xi= 0$ 
rather than $-14$ MeV, 
based on the assumption of $\Xi$ absorption in the nucleus\cite{kohno}. 
More recently, the data from the BNL-E906 experiment\cite{e906}, 
which investigated the ${^9 \text{Be}}(K^{-},K^{+})X$ reaction at 1.8 GeV$/c$, 
was reanalyzed using a phenomenological approach by T.Harada \textit{et al.}\cite{harada}. 
This analysis found an attractive $V_{0}^{\Xi} \approx -17 \pm 6 \ \text{MeV}$ 
with an absorption term $W_{0}^{\Xi} \sim 5 \ \text{MeV}$. 
However, it is unlikely that the line-shape of the $\Xi$ production spectrum 
is sensitive to the absorption strength. 

The KEK E224 experiment provided direct information on the scattering cross section 
of the $\Xi^-p$ interaction \cite{e224}.
An upper limit on the total cross section for $\Xi^-p$ elastic scattering was determined to be 24 mb
at a 90\% confidence level. Additionally, the total cross section for the $\Xi^-p\to\Lambda\Lambda$ reaction was measured to be $4.3^{+6.3}_{-2.7}$ mb. 
This measurement leads to an upper bound for the decay width of the $\Xi^-$, 
given by $\Gamma_\Xi< 3$ MeV. It is also noteworthy that the BESIII collaboration reported 
the cross section for the $\Xi^0n\to\Xi^-p$ process as 
$\sigma=7.4\pm 1.8({\rm stat.})\pm 1.5({\rm syst.})$ mb at $p_{\Xi^0}=0.818$ GeV$/c$, using the 
$\Xi^0+{}^9{\rm Be}\to\Xi^-+p+{}^8{\rm Be}$ in $e^+e^-$ collisions \cite{bes3}.

{\it 2. J-PARC E42 Experiment}  We measured cross sections for double-strangeness 
production processes in the $^{12} \text{C} (K^{-},K^{+})$ reaction using a separated 
1.82 GeV$/c$ beam at the K1.8 beam line 
of the Hadron Experimental Facility at J-PARC\cite{k18, k18_ref2}. The $K^{-}$ beam 
was directed onto a synthetic diamond target,  2 cm in height, 3 cm in width, and 2 cm in thickness with  a density of 3.223 g$/$cm$^3$,
resulting in the production of double-strangeness systems in the $(K^{-},K^{+})$ reaction by
tagging $K^+$ particles in the forward direction. 
Additionally, a polyethylene (CH$_2$) target was employed to collect reference data 
for calibration purposes. 
%
%%%%%%%%%%%%%%%%%%%%%%%%%%%%%%
\begin{figure}[!htb]
\centering
\includegraphics[width=0.7\textwidth]{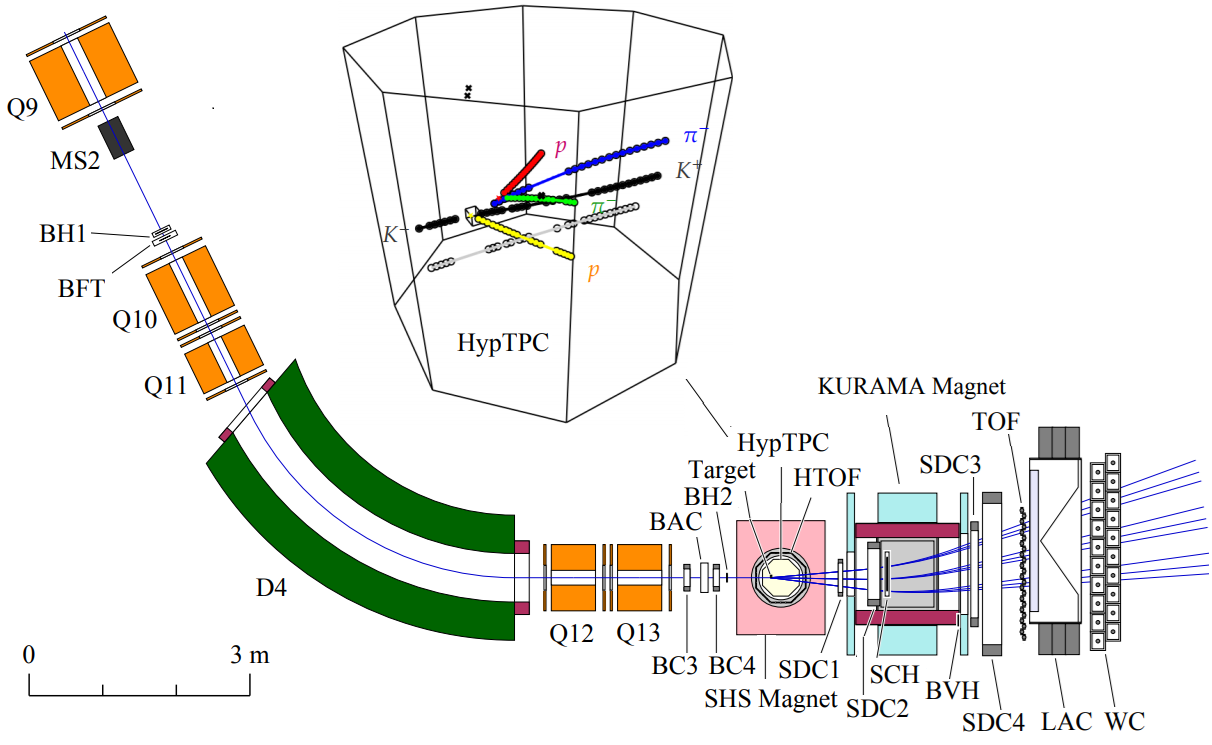}
\caption{\small The E42 detector configuration.
A typical event recorded by the HypTPC is displayed above.}
\label{fig:e42setup}
\end{figure}
%%%%%%%%%%%%%%%%%%%%%%%%%%%%%%

The J-PARC E42 detector system consists of three main components: 
the K1.8 beam-line spectrometer, which measures the incident $K^-$ beam; 
the Superconducting Hyperon Spectrometer (SHS), 
which is used to measure charged particles from the diamond target; 
and the forward KURAMA spectrometer, which 
detects the outgoing $K^+$ particles, as depicted in Fig. \ref{fig:e42setup}.

The SHS is a key feature of J-PARC E42 detector,  
providing  wide-angle coverage for detecting charged particles from the $(K^-,K^+)$ reaction.
The SHS features a superconducting magnet that operates 
at 1 T\cite{magnet}, a time-projection chamber (HypTPC) \cite{hyptpc}, 
and a time-of-flight (HTOF) array \cite{htof} surrounding the HypTPC. 
A diamond target is positioned 143 mm upstream from 
the geometric center of the HypTPC. The HypTPC has an octagonal prism-shaped drift
volume with 55 cm in drift length and a circular pad plane with concentric pad rows centered
at the target position.
The inner sector contains 10 radial pad rows, with pads 
of 9 mm in length and 2.1--2.7 mm in width.  
The outer sector, on the other hand, has 22 pad rows with 
pads that are 12.5 mm in length and have widths ranging from 2.3 mm to 2.4 mm.
This configuration results in a total of 5768 pads. 

The momenta of outgoing particles are measured using the KURAMA spectrometer, which
spans approximately 3.5 m in length. This spectrometer consists of
a dipole magnet with a field strength of 0.73 T$\cdot$m, 
two sets of drift chambers (SDC1 and SDC2) with five planes each located at the entrance of the magnet, and two additional drift chambers (SDC3 and SDC4) with four planes each at its exit.
It also includes large silica aerogel and water Cherenkov counters (LAC and WC), 
as well as two hodoscopes (SCH and TOF) for timing and triggering purposes. 

The combination of the Hyperon Spectrometer and the forward KURAMA spectrometer covers
a $K^+$ laboratory angle range of $\theta_{K^+} < 25^\circ$ 
and a momentum range of $p_{K^+} > 0.5$ GeV$/c$. The momentum resolution $(\Delta p/p)$
is 2.7\% (FWHM) at 1.3 GeV$/c$. Further details on the performance of the E42 detector 
will be provided in a separate technical article.

%%%%%%%%%%%%%%%%%%%%%%%%%%%%%%%%%%%%%%%

{\it 3. Data Analysis} 
The momentum of the incident particle was measured using hit information 
from a scintillating fiber hodoscope (BFT), 
which is located upstream of the K1.8 QQDQQ beam spectrometer, 
and two drift chambers (BC3 and BC4), placed downstream of the last beamline magnet. 
To suppress pion background, the time of flight was determined from signals 
between two scintillator hodoscopes (BH1 and BH2) and the beam aerogel counter (BAC).
The $K^-$ beam recontruction efficiency is 96.8\%.  

The momentum of the outgoing $K^+$ was measured 
using a combination of the SHS spectrometer and the KURAMA spectrometer. 
Hit information from the HypTPC and four drift chambers was utilized 
to reconstruct the trajectory of the outgoing $K^+$. 
The track parameters were obtained by propagating the tracks through 
two magnetic spectrometers using the fourth-order Runge-Kutta method. 
After applying goodness-of-fit, multiplicity, and vertex cuts, 
the mass of the outgoing particle was calculated based on the measured momentum, 
path length, and time-of-flight information between BH2 and the TOF. 
The tracking efficiency with detector components of the KURAMA spectrometer
was typically 94\%, while the $K^+$ track-finding efficiency in the HypTPC was 92\% on
average.  
%
%%%%%%%%%%%%%%%%%%%%%%%%%%%%%%
\begin{figure}[!htb]
\centering
\stackinset{c}{-0.1cm}{b}{-0.5cm}{(a)}{
\stackinset{l}{2.3cm}{b}{1.0cm}{$\pi^+$}{
\stackinset{l}{3.0cm}{b}{1.0cm}{$K^+$}{
\stackinset{r}{2.1cm}{b}{1.0cm}{$p$}{
\includegraphics[width=0.465\textwidth]{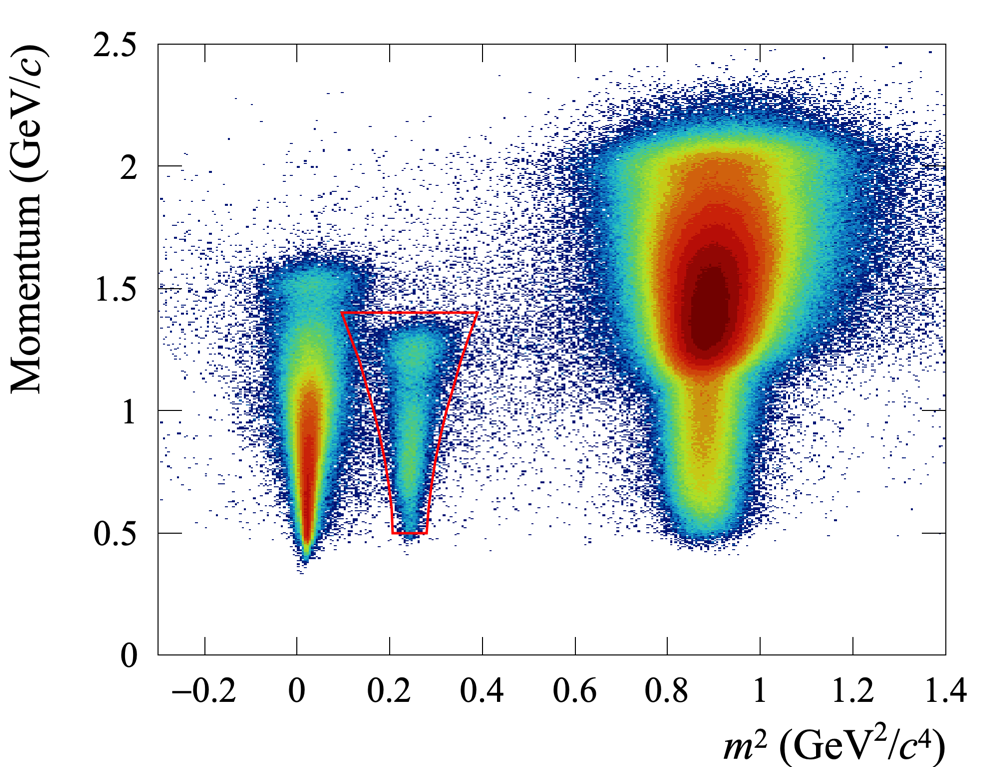}
}}}}
\stackinset{c}{-0.1cm}{b}{-0.5cm}{(b)}{
\includegraphics[width=0.465\textwidth]{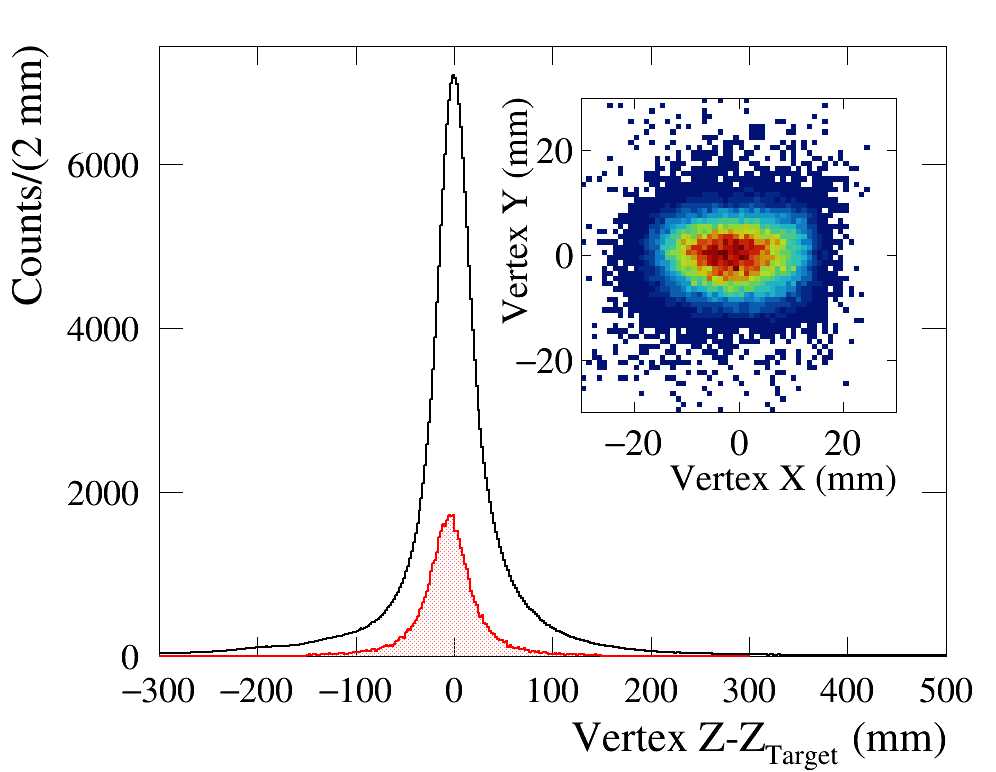}
}
\caption{\small (a) A scatter plot of reconstructed mass squared distribution for outgoing particles versus their momenta and
(b) vertex distribution in the $Z$ direction. 
The red shaded area indicates the vertex distribution 
for both $\Xi^-$ and $\Lambda\Lambda$ events. 
The inset displays the vertex distribution in the $XY$ plane.}
\label{fig:fig2}
\end{figure}
%%%%%%%%%%%%%%%%%%%%%%%%%%%%%%
%
A scatter plot displaying the reconstructed mass squared versus momentum for outgoing particles is presented in Fig. \ref{fig:fig2}(a). 
Kaons were identified by selecting a $3\sigma$ window of 
the reconstructed mass squared in conjunction with momentum, 
which is indicated as the red area of Fig. \ref{fig:fig2}(a). 
Using the reconstructed $K^-$ and $K^+$ tracks, 
we reconstructed the $(K^-, K^+)$ reaction vertex, where the two tracks 
intersect at their closest point.
The distributions of the two-track vertices are shown in Fig. \ref{fig:fig2}(b). 

We first selected events based on the requirement that 
the $Z$ vertex lies within a tolerance window of 150 mm from the target center. 
The two-track vertex resolution in the $Z$ direction 
ranges from 5 ($\theta^{\rm lab}_{K^+}=15^\circ$) 
to 20 mm  ($\theta^{\rm lab}_{K^+}=7^\circ$).
The shaded area in the distribution illustrates the vertex distribution 
for $\Xi^-$ production events. 
An inset displays the reconstructed vertex distribution on the $XY$ plane. 
With the addition of one more track, we can reconstruct a three-track vertex,
achieving resolutions of 3.7 mm in the $Z$ direction and 0.70 mm and 
0.45 mm in the $X$ and $Y$ directions, respectively. 

When a charged particle passes through the HypTPC drift volume, its trajectory
is recorded as hits on the pad plane and arrival timing of the pulses
giving the position along the drift direction.  
A hit is defined as an active pad that produces a signal above a certain threshold.  
Clusters are identified as groups of active pads that have at least two consecutive hits 
on the same pad row. The coordinates of a cluster are determined by calculating
the center of gravity of the hits.
Initial track finding is performed using the Hough-transform method.
A track must be associated with at least six clusters, 
which corresponds to a minimum track length of 5.4 cm.

The initial track parameters are determined through 
a reduced $\chi^2$ minimization of track residuals, 
with the assumption that a track follows a helical trajectory within the HypTPC. 
Track-associated clusters must fall within $5\sigma$ windows of the transverse 
and vertical spatial resolutions. 

The transverse spatial resolution is modeled 
as the quadratic sum of the intrinsic resolution and three additional terms that 
depend on the readout pad length, drift distance, and the track angle 
with respect to the pad row \cite{spatial_res}. 
The horizontal intrinsic resolutions are 387 $\mu$m for the long pads and 750 $\mu$m 
for the short pads located near the target. 
On average, the transverse spatial resolution is approximately 400 $\mu$m.
The vertical spatial resolution is 1.0 mm and shows minimal dependence on the drift length.

Further track reconstruction was performed using the least-squares 
Kalman filtering algorithm, which is implemented in GENFIT, 
a generic framework for track fitting \cite{genfit}. 
At this stage, energy loss corrections were applied due to interactions
with the material, particularly in the diamond target.
The particle identification in the HypTPC focuses on 
distinguishing between pions and protons, 
because the incoming $K^-$ beam and the outgoing $K^+$ tracks are identified 
using other spectrometer detectors. 
The separation between pions and protons is based on the energy loss information 
of charged tracks within the HypTPC gas volume. 
%
%%%%%%%%%%%%%%%%%%%%%%%%%%%%%%
\begin{figure}[!htb]
\centering
\stackinset{c}{-0.1cm}{b}{-0.5cm}{(a)}{
\includegraphics[width=0.625\textwidth]{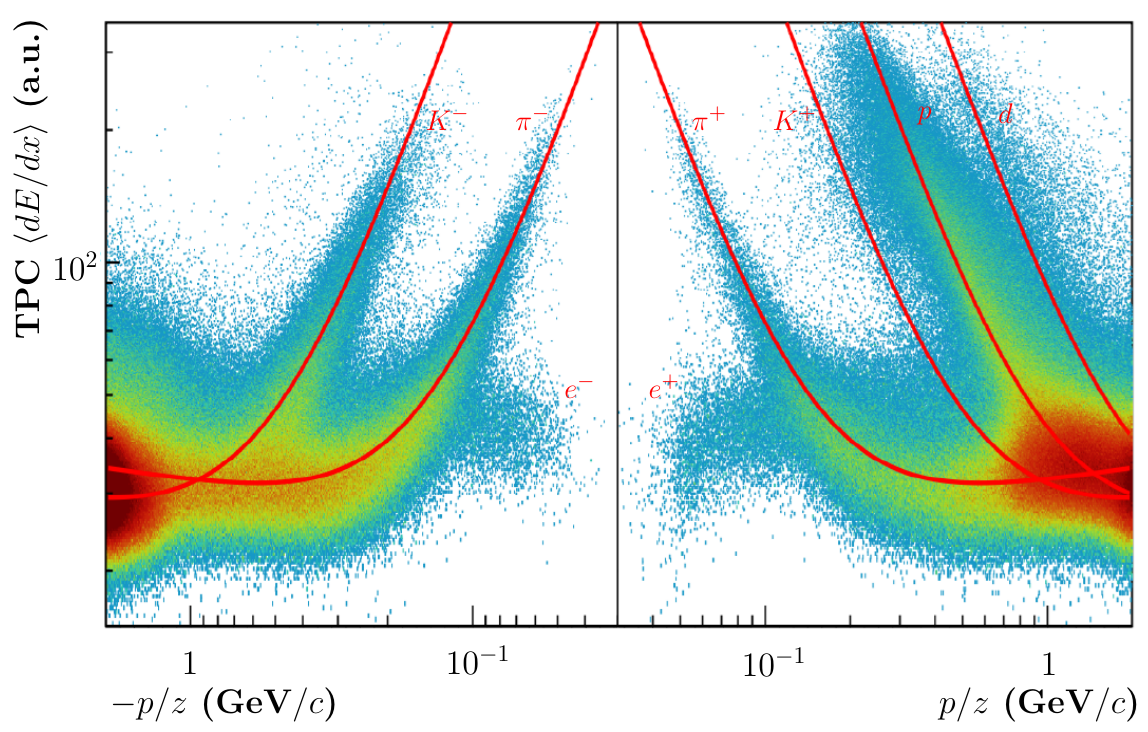}
}
\hskip-0.2cm
\stackinset{c}{-0.1cm}{b}{-0.5cm}{(b)}{
\stackinset{c}{-0.5cm}{t}{0.7cm}{$\Lambda$}{
\stackinset{c}{-0.5cm}{t}{3.4cm}{$\Xi^-$}{
\raisebox{0.25cm}{
\includegraphics[width=0.31\textwidth]{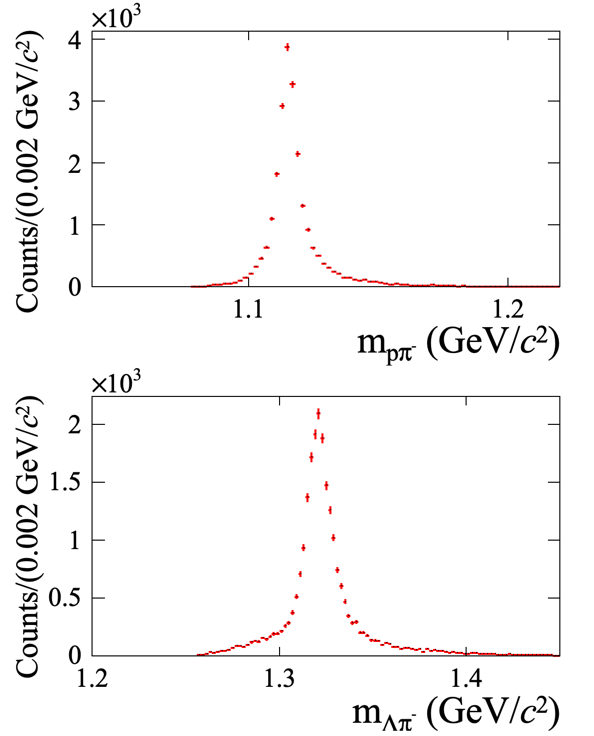}
}
}}}
\caption{\small (a) Specific energy loss in HypTPC versus reconstructed momentum/charge and (b) the invariant mass distributions for the $p\pi^-$ and $\Lambda\pi^-$ systems.}
\label{fig:fig3}
\end{figure}
%%%%%%%%%%%%%%%%%%%%%%%%%%%%%%

Particle identification is based on measuring the specific energy loss
($\langle dE/dx\rangle$) of the track in the HypTPC. To minimize fluctuations
from the Landau tail, $\langle dE/dx\rangle$ is defined as the truncated mean
of the 80\% lowest HypTPC clusters associated with a track. 
The energy loss is primarily determined by the charge and rest mass of the particle within a given momentum bin.

For separating protons from pions ($p/\pi$ separation),
 we use a separation criterion defined as: 
$\langle dE/dx \rangle_{\pi} + 0.5 S_{p/\pi}\sigma_{dE/dx_\pi}$, 
 where the $p/\pi$ separation power ($S_{p/\pi}$) is expressed as:
$S_{p/\pi} = |\langle dE/dx \rangle_p - \langle dE/dx \rangle_{\pi}|/\left[\left( \sigma_{dE/dx_p} + \sigma_{dE/dx_\pi} \right)/2\right]$.
In this formula, $\langle dE/dx \rangle_p$ and $\langle dE/dx \rangle_{\pi}$ represent 
the mean energy losses for protons and pions, respectively, while $\sigma_{dE/dx_p}$ and 
$\sigma_{dE/dx_\pi}$ are the corresponding standard deviations of these energy losses. 
The momentum threshold for achieving a separation power of $3\sigma$ 
between protons and pions is 0.48 GeV$/c$. In this study, 
the pion has a sufficiently low momentum, allowing for excellent $p/\pi$ separation. 

In the $\Lambda\to p\pi^-$ decay, we require that
the pion and proton tracks interact within a distance of closest approach (DCA) of
less than 1 cm. The reconstructed mass of the $\Lambda$ candidate should fall 
within a window of $\pm$100 MeV$/c^2$. We then perform a mass-constraint fit based on
the $\Lambda\to p\pi^-$ decay hypothesis.
For the reconstruction of the $\Xi^-$ decay, we require that the DCA between the $\Lambda$
and an additional $\pi^-$ track is less than 1.5 cm and the reconstructed mass of the $\Xi^-$
must also be within a window of $\pm$100 MeV$/c^2$. 

Additionally, we ensure that the particle tracks originating 
from the $(K^-, K^+)$ production vertex intersect at the center of the target. 
The closest distance must be less than 2.5 cm, while 
the vertical distance should not exceed 2.0 cm.
This additional condition helps to minimize incorrect pairings.
The reconstructed mass distributions for $\Lambda$ and $\Xi^-$  are
displayed in Fig. \ref{fig:fig3}.

For $\Lambda\Lambda$ events, we fully reconstruct the two $\Lambda$ decays
using two $p\pi^-$ pairs. The reconstruction of the two $\Lambda$ particles is performed by
iterating over two possible $p\pi^-$ combinations. A vertex fit is applied 
to each combination, ensuring that the $p\pi^-$ particles originate from 
common decay vertices. The point at which the $\Lambda$ decays into a proton and a pion
is referred to as the decay vertex, while the point from which the $\Lambda$ originated is
known as the production vertex. The production vertex is determined by reconstructing
the $(K^-, K^+)$ reaction vertex. We require that the mass of the $\Lambda$ candidates falls within
a window of $\pm$100 MeV$/c^2$. All identified $\Lambda$ candidates are retained for further analysis.

A mass-constraint fit is performed under the hypothesis of $\Lambda$ or $\Xi^-$ decay. 
Event selection for $\Lambda\Lambda$ production involves 
differentiating between two $\Lambda$ decays and $\Xi^-p$ emission 
from the $(K^-, K^+)$ reaction vertex. 
Both cases lead to the same final state, which consists of two protons and two $\pi^-$ tracks. 
Consequently, some reconstructed events may satisfy the selection criteria 
that require two valid $\Lambda$ decays for $\Lambda\Lambda$ production 
and one valid $\Xi^-$ for the $\Xi^-p$ emission event as well.

The event selection process is based on Bayes' theorem, 
which employs prior probability distributions to generate posterior probabilities. 
The probability density function (PDF) for $\Lambda\Lambda$ production is created 
using simulated signal events for $\Lambda\Lambda$. 
In contrast, the PDF for $\Xi^-p$ emission is derived from $\Xi^-p$ events 
that have been misidentified as $\Lambda\Lambda$ events. 
The prior probability, denoted as $P_{\Lambda\Lambda}^{\text{prior}}$, 
for the $\Lambda\Lambda$ reaction is obtained from the template fit 
to the simulated distributions of invariant mass and lifetime 
for both $\Lambda$ and $\Xi^-$ decays.

Using Bayes' theorem, the posterior probability of the occurrence of 
the $\Lambda\Lambda$ reaction, 
given the observed kinematic variables $(x_1, x_2)$ is expressed as:
\begin{equation}
    P(\Lambda\Lambda \mid x_1, x_2) = p_{\Lambda\Lambda} (x_1, x_2) P_{\Lambda\Lambda}^{\text{prior}} /
    \left[ p_{\Lambda\Lambda} (x_1, x_2) P_{\Lambda\Lambda}^{\text{prior}} + p_{\Xi^- p} (x_1, x_2) (1 - P_{\Lambda\Lambda}^{\text{prior}}) \right],
\end{equation}
where $p_{\Lambda\Lambda} (x_1, x_2)$ and $p_{\Xi^- p} (x_1, x_2)$ represent 
the probability density functions for each reaction. 
The kinematic variables $x_1$ and $x_2$ correspond to 
the reconstructed $\Lambda$ invariant mass 
and the reconstructed $\Lambda$ lifetime, respectively.

We require both $\Lambda$ candidates to satisfy $P_{\Lambda\Lambda} > 0.5$ for the $\Lambda\Lambda$ event selection. 
The selection efficiency was determined to be 85.6\% using the Geant4-based E42 simulation software.  
This efficiency was also incorporated into the overall experimental efficiency. 
The difference in efficiency when applying the requirement to only one of the two $\Lambda$ candidates was considered 
as a systematic uncertainty in the event selection process. This uncertainty is approximately 10\% of the total yield.

{\it 4. Experimental Results}
Data collected with a polyethylene (CH$_2$) target was used to 
investigate the missing-mass resolution. 
The missing-mass spectrum for the $(K^-, K^+)$ reaction on hydrogen in the CH$_2$ target 
reveals distinct peaks associated with the $\Xi^-$ and $\Xi(1535)^-$ production, 
along with a contribution from nonresonant $\Xi\pi$ production.
The background stemming from reactions 
on carbon has been subtracted using a normalized data sample obtained 
from the diamond target.
The missing mass resolution for the CH$_2$ target was measured to be $10.4\pm 0.5$ MeV$/c^2$. 
This resolution was then used to estimate the missing mass resolution for carbon, 
which was found to be approximately 15 MeV$/c^2$.

The missing mass for the $(K^-, K^+)$ reaction on carbon can be alternatively expressed 
in terms of the $\Xi^-$ excitation energy ($E_\Xi$) \cite{harada2}, which is defined as
the missing mass $M_X$ minus the combined mass of the $\Xi^-$ and $^{11}$B ground state.
This definition connects the negative binding energy of $\Xi^-$ to the its energy 
above the free $\Xi^-$ emission threshold ($E_\Xi=-B_\Xi$).
 
The double differential cross section was obtained using the following formula:
\begin{equation}
    \left\langle\frac{d^2 \sigma}{d\Omega dE}\right\rangle = \frac{A}{N_A (\rho x)} \frac{N_{KK}}{N_{\text{beam}} \Delta \Omega_{\theta_1{\textrm -}\theta_2} \Delta E \varepsilon}
\end{equation}
where $A$ represents the normalization factor, 
$N_A$ is Avogadro's number, $\rho$ and $x$ denote the target density and thickness, respectively, 
$N_{KK}$ is the number of detected $(K^-,K^+)$ events, 
$N_{\text{beam}}$ is the total number of incident $K^-$ beams, 
$\Delta \Omega_{\theta_1 - \theta_2}$ represents 
the solid angle acceptance within the scattering angle range from $\theta_1$ to $\theta_2$, 
$\Delta E$ is the bin width of $E_\Xi$, and $\varepsilon$ denotes the total experimental efficiency.

To assess the reconstruction efficiency for $\Lambda$ and $\Xi^-$ decays in the HypTPC, 
the HypTPC was integrated into the E42 Geant4-based simulation software, 
which provides a detailed representation of the drift volume, including a 
realistic model of the readout pad plane. 
The geometrical acceptance was defined as a function of the reconstructed $\Lambda$ momentum and polar angle.
To obtain the acceptance, we averaged over the azimuthal angle $\phi$.
The estimated reconstruction efficiency for $\Xi^-$ decays is approximately 65\% and for $\Lambda\Lambda$ pairs, about 50\% in the range of $E_{\Xi} = 100$–$150$ MeV. 

Systematic uncertainties in cross sections arise from fluctuations in beam tracking efficiencies 
and data acquisition efficiencies across different runs, which accounts 
for a variation of 0.1\%. 
Addtionally, uncertainties due to finite momentum resolution affect 
the $K^+$ survival ratio and the estimation of $\pi/p$ background, 
leading to uncertainties of 0.2\% and 0.003\%, respectively. 
Another source of systematic uncertainty stems from the acceptance correction process, 
which yields estimates of 8.8\% to 20.6\% for the $\Lambda\Lambda$ cross section, 
while remaining below 1\% for the $\Xi^-$ cross section.

\begin{figure}[!htb]
\centering
\stackinset{c}{-0.1cm}{b}{-0.5cm}{(a)}{
\includegraphics[width=0.47\textwidth]{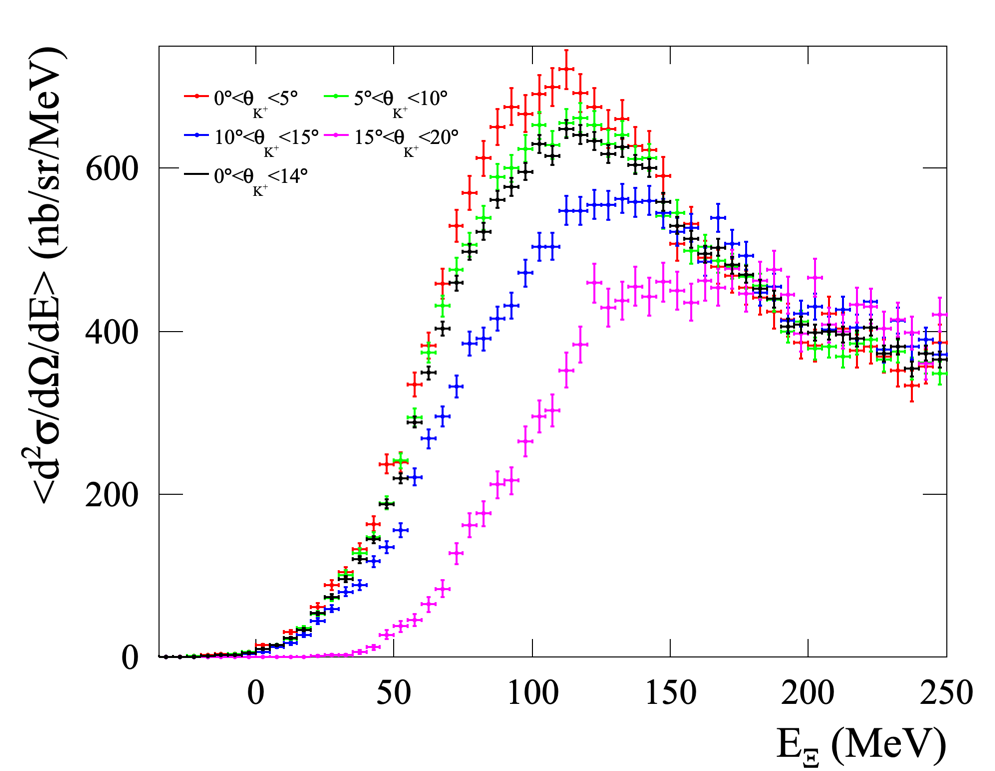}
}
\stackinset{c}{-0.1cm}{b}{-0.5cm}{(b)}{
\includegraphics[width=0.47\textwidth]{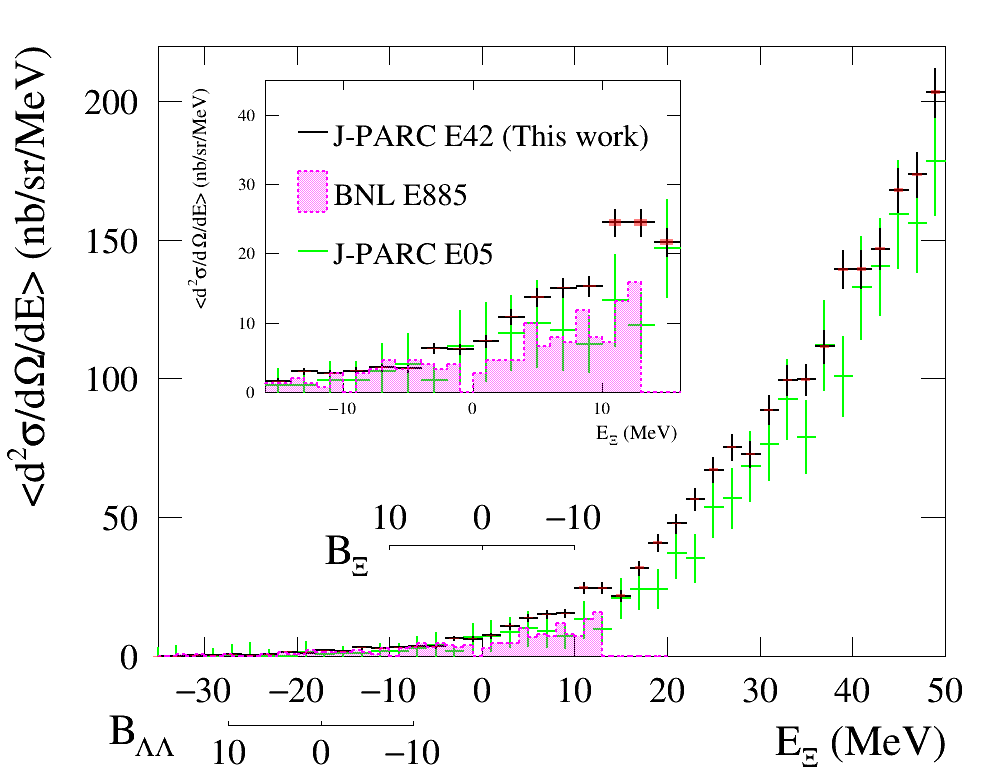}
}
\caption{\small (a) Differential cross sections for the inclusive $^{12}$C$(K^-,K^+)X$ reaction 
averaged over different $K^+$
lab scattering angles at 1.8 GeV$/c$ and (b) the cross section averaged over the region 
$0^\circ<\theta^{\rm lab}_{K^+}<14^\circ$ near $E_\Xi=0$. 
Pink shaded area and green crosses represent data from earlier measurement, BNL E885 \cite{e885} and the J-PARC E05 \cite{e05}. Systematic uncertainties are indicated by red boxes. Binding energies for
the $\Xi^-$ (B$_\Xi$) and $\Lambda\Lambda$ (B$_{\Lambda\Lambda}$) are presented on separate, overlaid axes.
}
\label{fig:fig4}
\end{figure}

Moreover, when the geometrical acceptance falls below 5\%, applying the acceptance correction 
leads to excessive fluctuations. To address this issue, events in this region were excluded from the analysis, 
resulting in the removal of approximately 5\% of all $\Lambda$ candidates. 
For $E_{\Xi}$ in the range of 100–150 MeV, the systematic uncertainty in the $\Lambda\Lambda$ yield 
is approximately 20\% of the total yield, while for $\Xi^-$, it remains below 1\%. In Fig.~\ref{fig:fig4}(b),
Fig.~\ref{fig:fig6}(a), and Fig.~\ref{fig:fig7}(b), systematic uncertainties are represented by red boxes.

%%%%%%%%%%%%%%%%%%%%%%%%%%%%%%
%%%%%%%%%%%%%%%%%%%%%%%%%%%%%%

The differential cross sections for the inclusive $^{12}$C$(K^-,K^+)$ reaction
were measured as a function of $E_\Xi$, 
averaging over different $K^+$ scattering angle regions in Fig. \ref{fig:fig4}(a).
The results, shown in Fig. \ref{fig:fig4}(b), include data
averaged over the angle of $0^\circ$ to $14^\circ$ and are compared
with the previous measurements near $E_\Xi=0$.
The findings demonstrate good agreement with
earlier measurements from J-PARC E05~\cite{e05} 
and BNL E885~\cite{e885}. 
%
%%%%%%%%%%%%%%%%%%%%%%%%%%%%%%
\begin{figure}[!htb]
\centering
\stackinset{c}{-0.1cm}{b}{-0.5cm}{(a)}{
\includegraphics[width=0.47\textwidth]{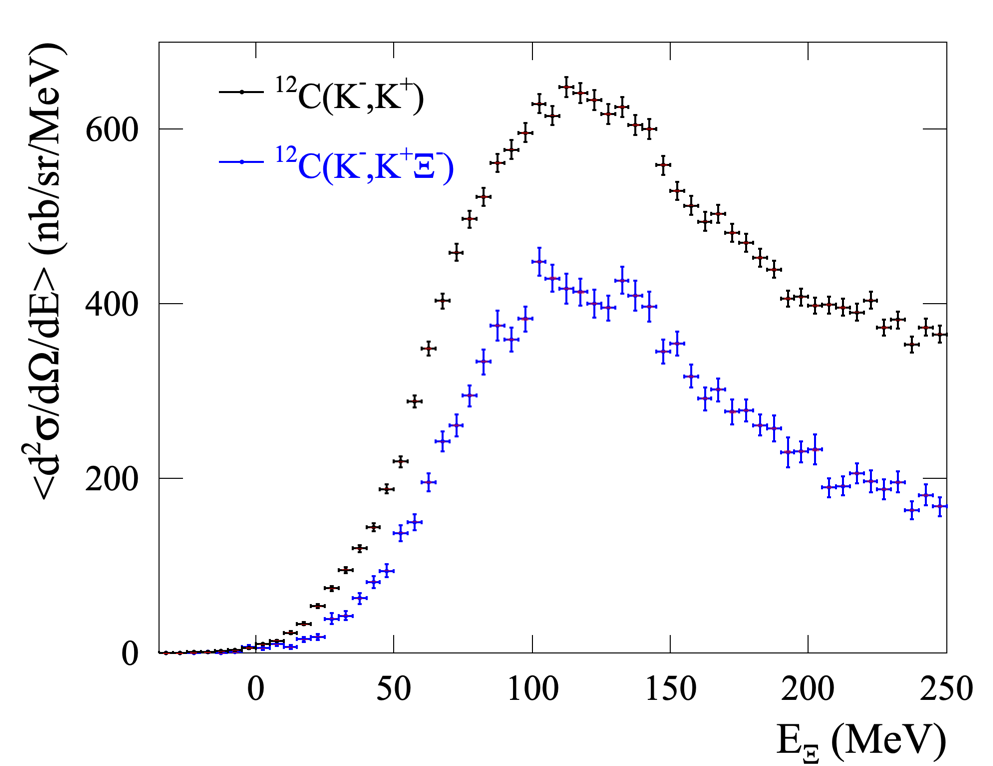}
}
\stackinset{c}{-0.1cm}{b}{-0.5cm}{(b)}{
\includegraphics[width=0.47\textwidth]{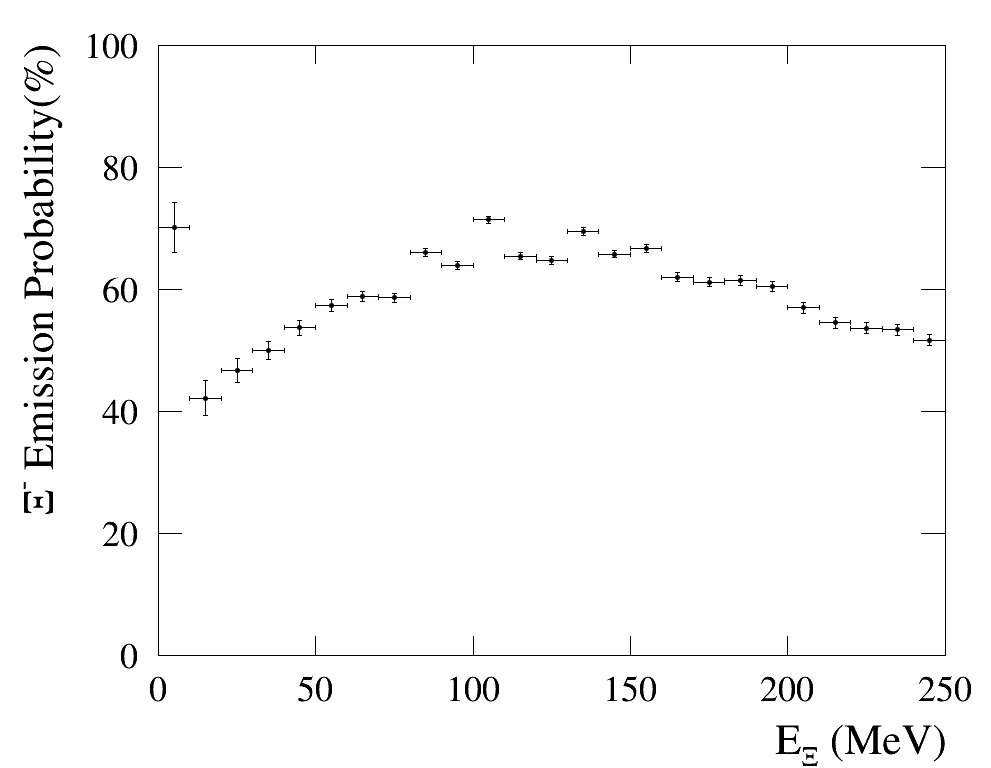}
}
\caption{\small (a) Differential cross sections for the $^{12}$C$(K^-, K^+\Xi^-)$ reaction
in comparison with those for the inclusive $^{12}$C$(K^-, K^+)$ reaction, averaged over
the $K^+$ angle of $0^\circ$ to $14^\circ$. Systematic uncertainties are indicated by red boxes. 
(b) The emission probability of $\Xi^-$ from $^{12}$C with respect to 
the $\Xi^-$ excitation energy $(E_\Xi)$. }
\label{fig:fig5}
\end{figure}
%%%%%%%%%%%%%%%%%%%%%%%%%%%%%%
%
The differential cross sections for the 
$^{12}$C$(K^-, K^+\Xi^-)$ reaction, averaging over the $K^+$ scattering angle of
$0^\circ$ to $14^\circ$ was measured and compared with those for the
inclusive $^{12}$C$(K^-, K^+)$ reaction, as shown in Fig. \ref{fig:fig5}(a).
The bump structure near $E_{\rm ext}=120$ MeV appears in both spectra for
$\Xi^-$ production and inclusive reactions. This similar shape in both cases 
demonstrates that the bump structure is dominated by $\Xi^-$ production from carbon. 

The $\Xi^-$ emission probability $(P^{\rm emis}_\Xi)$ is defined as
the cross section ratio of the $\Xi^-$ emission to the $\Xi^-$ production processes. 
The missing fraction $1-P_\Xi^{\rm emis}$ accounts for the inelastic channels, 
while $P_\Xi^{\rm emis}$ pertains to both the quasi-free $\Xi^-$ production and 
the $\Xi^-p$ elastic scattering channel.
The inclusive $(K^-, K^+)$ reaction may involve non-$\Xi^-$ production processes, 
including two-step processes involving intermediate mesons.
For instance, the $(K^-, \pi^0)$ reaction could be 
followed by the $(\pi^0, K^+)$ reaction, 
while the $(K^-, \bar{K}^0)$ reaction might be followed by the $(K^0, K^+)$ reaction. 
Although the latter could contributes to the broad range of $E_\Xi$ spectrum, 
our data show nearly zero events in the deeply bound region. 
This finding is consistent with the results from the J-PARC E05 experiment, 
indicating that contributions from two-step process involving 
$(K^-, \bar{K}^0)$ and $(K^0, K^+)$ reactions are negligible. 
In contrast, the meson-induced two-step processes, 
such as the $(K^-,\pi^0)$ reaction 
followed by the $(\pi^0, K^+)$ reaction could still contribute to the inclusive $(K^-, K^+)$ reaction. 
Based on the result of this work, we have subtracted the contribution 
from the two-step processes involving intermediate mesons 
from the inclusive $(K^-, K^+)$ reaction cross section 
in order to estimate the $\Xi^-$ emission probability $P_\Xi^{\rm emis}$, 
as illustrated in Fig. \ref{fig:fig5}(b). 
This probability peaks at around 70\% 
between $E_\Xi=100$ and 150 MeV, with an average value of 60\% when considering
the range from 0 to 200 MeV.

%%%%%%%%%%%%%%%%%%%%%%%%%%%%%%
\begin{figure}[!htb]
\centering
\stackinset{c}{-0.1cm}{b}{-0.5cm}{(a)}{
\includegraphics[width=0.47\textwidth]{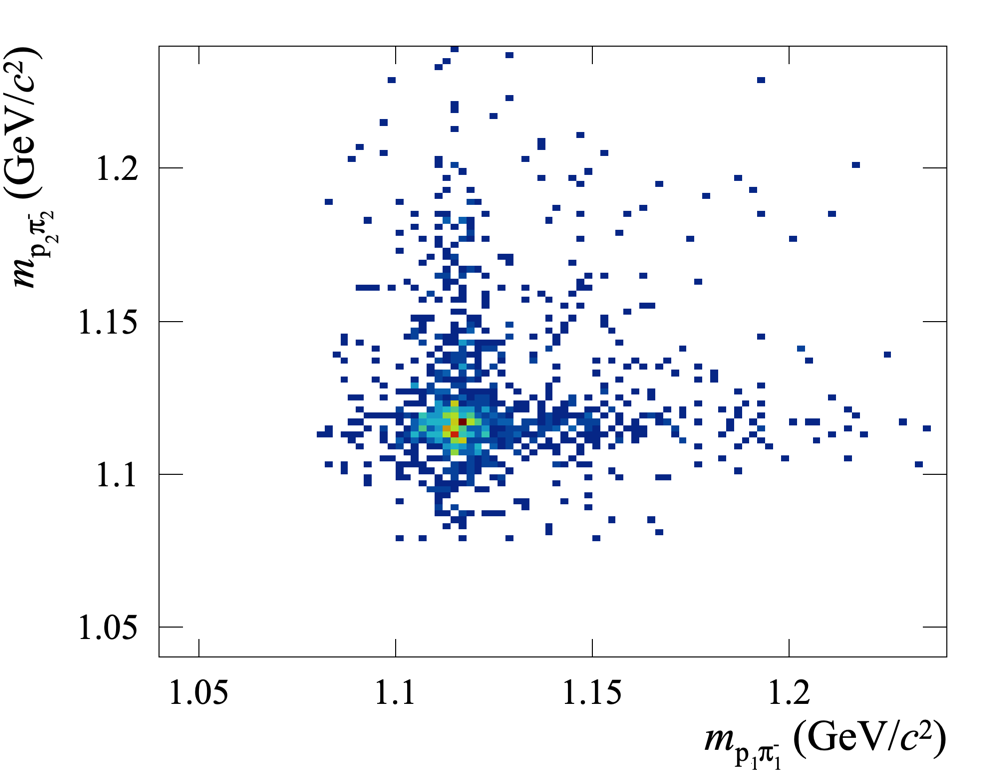}
}
\stackinset{c}{-0.1cm}{b}{-0.5cm}{(b)}{
\includegraphics[width=0.47\textwidth]{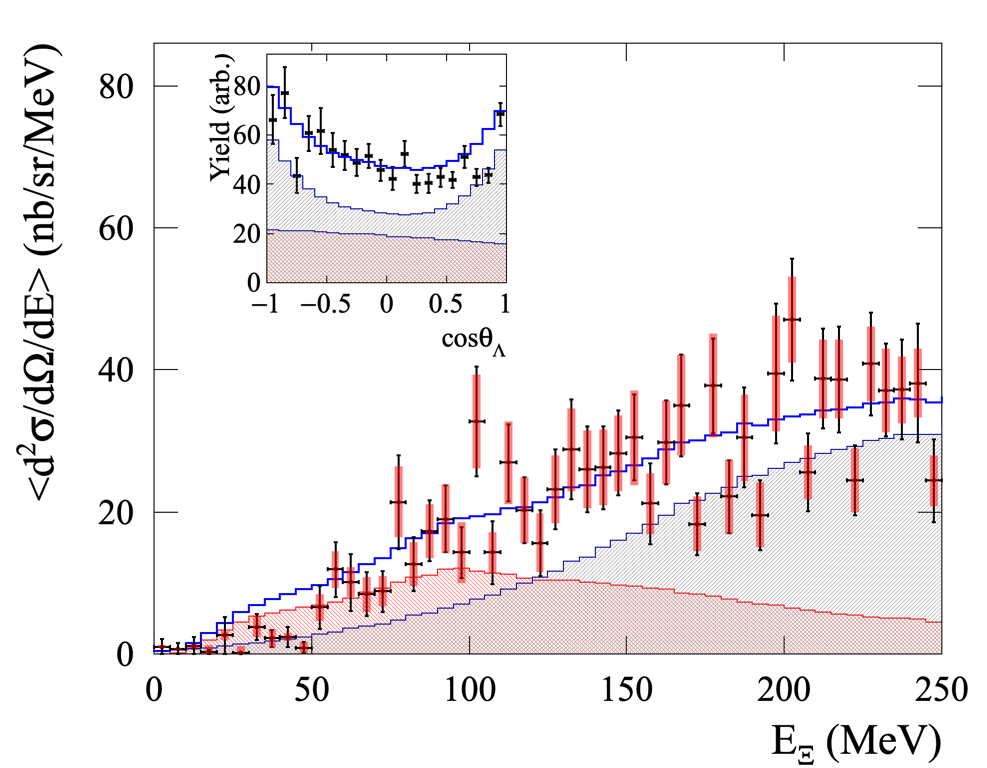}
}
\caption{\small (a) A scatter plot of the invariant masses for $(p\pi^-)_1$ and $(p\pi^-)_2$ 
and (b) differential cross sections for the $^{12}$C$(K^-,K^+\Lambda\Lambda)$ 
reaction. The results of intranuclear cascade calculations are compared, 
which account for the $\Xi^-p\to\Lambda\Lambda$ reaction (red shaded area) and two-step processes (gray shaded area) involving $\pi^0$, $\eta$, $\omega$ and $\eta^\prime$. 
The blue solid line corresponds to the sum of these contributions.
The inset displays 
the emission angle distribution of $\Lambda$ in the center-of-mass frame 
of the $\Xi^-p$ system, where the $\Xi^-$ momentum is given by
$\vec{p}_{K^-}-\vec{p}_{K^+}$. Systematic uncertainties are indicated by red boxes.}
\label{fig:fig6}
\end{figure}
%%%%%%%%%%%%%%%%%%%%%%%%%%%%%%
%
The reconstructed masses of two $p\pi^-$ pairs are shown in Fig. \ref{fig:fig6}(a). 
The differential cross section for the $^{12}$C$(K^-, K^+\Lambda\Lambda)$ reaction 
is depicted in Fig. \ref{fig:fig6}(b), 
showing a smooth increase as $E_\Xi$ rises. 
The integrated cross section in the energy region from 0 to 250 MeV is 
measured to be $5.13\pm 0.19$(stat.)$\pm 0.16$(syst.) $\mu$b/sr, which 
corresponds to 5\% of the inclusive $(K^-, K^+)$ cross section for carbon. 

The production of $\Lambda\Lambda$ is associated with the $\Xi^-p\to\Lambda\Lambda$
conversion process. Additionally, there are two-step processes involved,  
such as $K^-p\to\Lambda\pi^0$, followed by
$\pi^0p\to\Lambda K^+$. 
The $\pi^0$ can be replaced by other intermediate mesons,
including $\eta$, $\omega$, and $\eta^\prime$.  
To distinguish between these different processes, 
the production angle of $\Lambda$ was measured in the center-of-mass frame 
of the $\Xi^-p$ system.
The $\Xi^-$ is reconstructed using
the momentum transfer from the $(K^-, K^+)$ reaction with its mass 
($\vec{p}_\Xi=\vec{p}_{K^-}-\vec{p}_{K^+}$), while the 
target proton is assumed to be at rest.

Furthermore, the cross sections for $\Lambda\Lambda$ production 
are compared with calculations from
the intranuclear cascade model \cite{inc}, which describes secondary processes 
involving both $\Xi^-$-induced and intermediate meson-induced reactions, 
as illustrated in Fig. \ref{fig:fig6}(b).

{\it 5. Discussion}
The cross sections for the production of $\Xi^-$ and $\Lambda\Lambda$ provide additional
insight into the $\Xi^-p$ inelastic cross sections, based on the relative yield for $\Xi^-$
and $\Lambda\Lambda$ to the inclusive $(K^-, K^+)$ reaction events.  
To estimate the cross sections for $\Xi^-p$ inelastic scattering and $\Xi^-p\to\Lambda\Lambda$
reaction, we used the $\Xi^-$ emission probability and the measured $\Lambda\Lambda$
production cross sections
using a classical approach in the eikonal approximation.  

The effective number of the $\Xi N$ process is calculated as
\begin{eqnarray}
N_{\Xi N}=\int^\infty_0 2\pi bdb\int^{+\infty}_{-\infty} dz\rho(\sqrt{b^2+z^2})
F(b,z) \left[1-\exp\left\{\bar{\sigma}_{\Xi N}\int^{+\infty}_z dz^\prime\rho(\sqrt{b^2+z^{\prime 2}}
\right\}\right]  
\end{eqnarray}
where the initial and final state interactions are taken into accout using a factor
$F(b,z)=\exp\left[-\bar{\sigma}_{K^-N}\int^z_{-\infty} dz^\prime\rho(\sqrt{b^2+z^{\prime 2}}) 
-\bar{\sigma}_{K^+N}\int^{+\infty}_z dz^\prime\rho(\sqrt{b^2+z^{\prime 2}}\right].$
The nuclear density is modeled with a Fermi distribution given by: 
$\rho(r)=(1-wr^2/r_c^2)/(1+\exp(r-r_c/a))$, where $w=0.149$, $a=0.5224$, and $r_c=2.355$ fm.
The mass number $A$ is calculated as: $A=\int d\vec{r}\rho(\vec{r})$. 
The averaged cross sections are taken to be
$\bar{\sigma}(K^-N)= 28.9$ mb and $\bar{\sigma}(K^+N)= 19.4$ mb, respectively. The effective
number of the $(K^-, K^+)$ reaction events is given by:
$N_{KK}=\int^\infty_0 2\pi bdb\int^{+\infty}_{-\infty} dz\rho(\sqrt{b^2+z^2}) F(b,z)$. 
%
%%%%%%%%%%%%%%%%%%%%%%%%%%%%%%
\begin{figure}[!htb]
\centering
\stackinset{c}{-0.1cm}{b}{-0.5cm}{(a)}{
\includegraphics[width=0.47\textwidth]{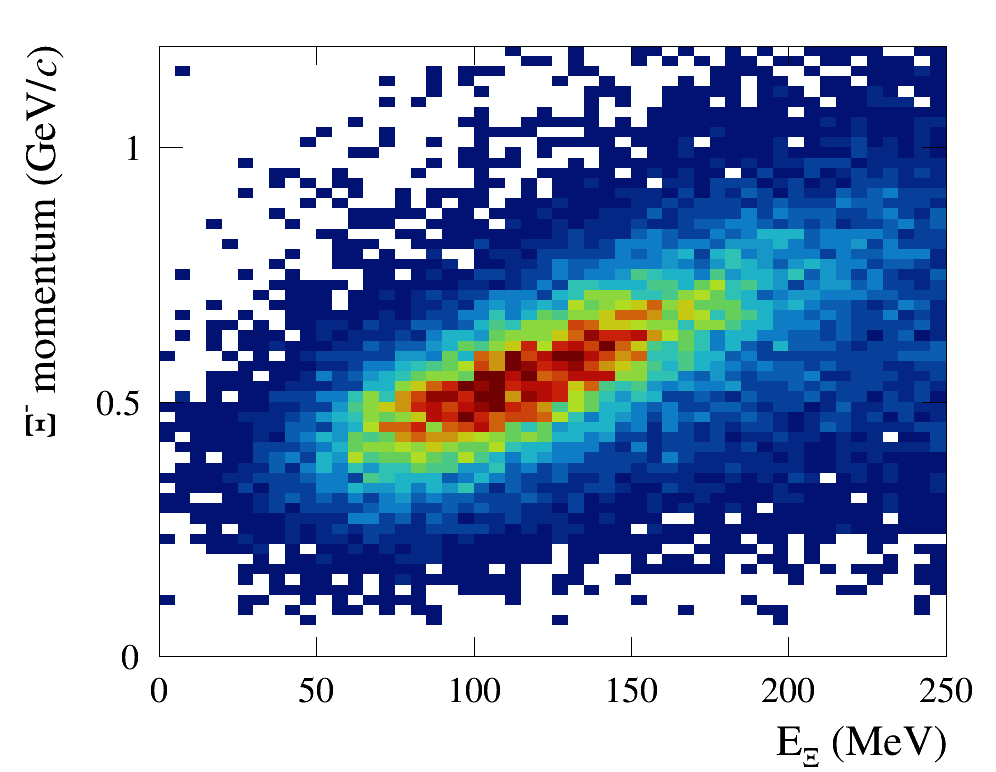}
}
\stackinset{c}{-0.1cm}{b}{-0.5cm}{(b)}{
\includegraphics[width=0.47\textwidth]{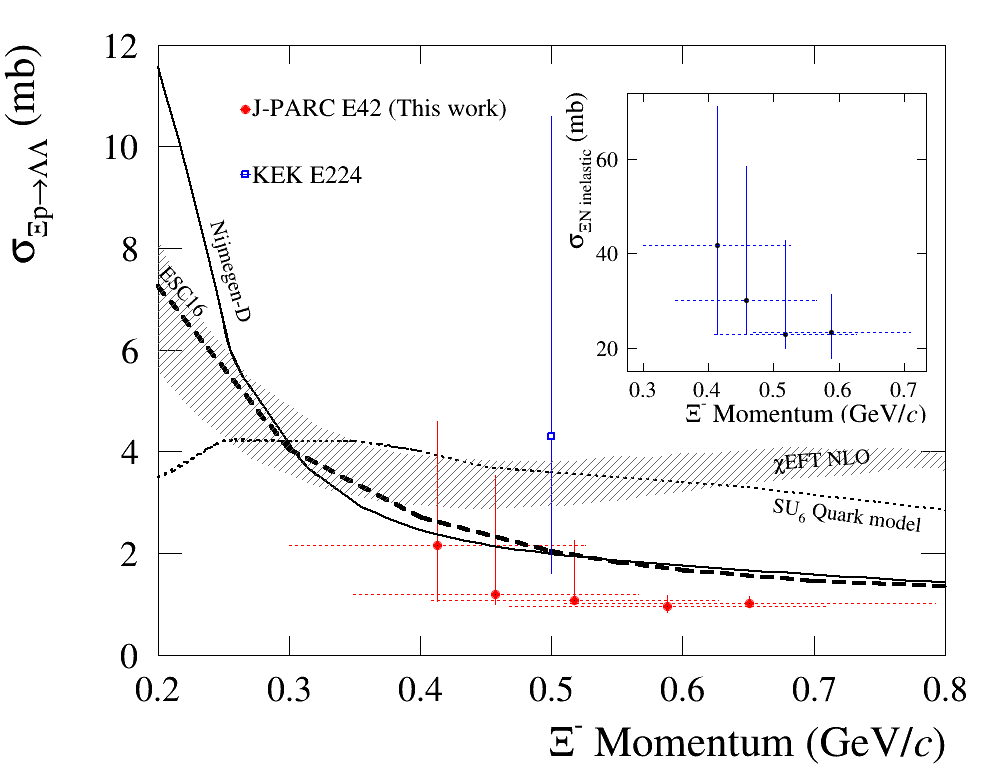}
}
\caption{\small (a) A scatter plot of the correlation between the $\Xi^-$ excitation energy and its
momentum and (b) the total cross sections for the $\Xi^-p\to\Lambda\Lambda$ reaction
as a function of $\Xi^-$ momentum.
The $\Xi^-$ momentum regions in adjacent energy bins overlap, 
which is represented by the horizontal dashed bars. 
The blue crosses represent data from the previous KEK-E224 experiment, while the curves
depict theoretical predictions from the Nijimegen-D model \cite{njd}, 
the ESC16 model \cite{esc16},
the SU$_6$ quark model \cite{qm}, 
and the chiral effective field theory up to next-to-leading order ($\chi$EFT NLO) \cite{eft}. 
The inset displays the total cross sections for $\Xi^-p$ inelastic scattering. }
\label{fig:fig7}
\end{figure}
%%%%%%%%%%%%%%%%%%%%%%%%%%%%%%
%
The $\Xi^-$ excitation energy ($E_\Xi$) is correlated 
with its $\Xi^-$ momentum reconstructed from its decay products, 
as shown in Fig. \ref{fig:fig7}(a). 
This study examines five energy bins ranging from $E_\Xi=30$ to 150 MeV, in 30 MeV intervals, 
since the non-resonant $\Xi\pi$ production may occur above $E_\Xi=150$ MeV. 
The $\Xi^-$ momentum regions in adjacent energy bins overlap, 
which is represented by the horizontal dashed bars in Fig. \ref{fig:fig7}(b). 
The systematic uncertainty in the cross section primarily arises from the
momentum spread of the $\Xi^-$ particles in a specific $E_\Xi$ bin. This uncertainty
is estimated as the maximum difference between the extrapolated values of the 
cross section at $\pm 1\sigma$ of the nominal $\Xi^-$ momentum.
From this analysis, the $\Xi^-p$ inelastic cross sections are estimated to range from
42 to 23 mb in the $\Xi^-$ momentum region of 0.4 to 0.6 GeV$/c$, as shown in the inset of
Fig. \ref{fig:fig7}(b). The average value is 
$22^{+14}_{-7}$ mb for the $\Xi^-$ momentum range of 0.5 to 0.6 GeV$/c$, 
Furthermore, the estimated total cross sections for the $\Xi^-p\to\Lambda\Lambda$ reaction are
presented in Fig. \ref{fig:fig7}(b), which are 
$2.2^{+2.5}_{-1.1}$ mb at 0.41 GeV$/c$, $1.2^{+2.4}_{-0.2}$ mb at 0.46 GeV$/c$, $1.1^{+1.2}_{-0.1}$ mb 
at 0.52 GeV$/c$, $1.0^{+0.2}_{-0.2}$ mb at 0.59 GeV$/c$, and 
$1.1^{+0.3}_{-0.03}$ mb at 0.65 GeV$/c$.
The average cross section in the region from 0.5 to 0.6 
GeV$/c$ is estimated to be $1.0^{+1.3}_{-0.8}$ mb, 
assuming the $\Lambda$ sticking probability of 0.15 \cite{Lstick}. 
This measurement result is likely to support the prediction by the Nijmegen-D model \cite{njd}
and the ESC16 model \cite{esc16}, 
rather than the SU$_6$ quark model \cite{qm} and 
the chiral effective field theory up to next-to-leading order \cite{eft}.

The total cross section for the $\Xi^-p\to\Lambda\Lambda$ reaction could impose a constraint on
the upper bound of the decay width for a $\Xi^-$ particle state in nuclear
matter, which can be calculated using the relation: $\Gamma_{\Xi}\approx
(v\sigma)_{\Xi^-p\to\Lambda\Lambda}\cdot(\rho_0/2)$, where $\rho_0/2\sim 0.08$ fm$^{-3}$
represents the proton density, and $v$ is the $\Xi^-p$ relative velocity \cite{dover}. 
Based on our estimation of the cross section for the $\Xi^-p\to\Lambda\Lambda$
reaction, which is $1.0$ mb, 
we find that the upper bound of the $\Xi^-$ decay width is approximately 
$\Gamma_{\Xi}\sim$ 0.6 MeV.
This narrow width of the $\Xi^-$ particle state supports the model based on the assumption of
$\Gamma_{\Xi}=0$ to fit the $E_\Xi$ spectrum in the previous experiment, 
J-PARC E05 \cite{e05}.
Moreover, this result suggests that the J-PARC E70 \cite{e70} has significant potential for
measuring a narrow $\Xi$-hypernuclear peak structure.

{\it 6. Conclusions}
We have measured cross sections for the $^{12}$C$(K^-, K^+\Xi^-)$ and 
$^{12}$C$(K^-, K^+\Lambda\Lambda)$ 
reactions at an incident beam momentum 1.8 GeV$/c$, 
with respect to $\Xi^-$ excitation energy.
The relative cross section for $^{12}$C$(K^-, K^+\Xi^-)$ reaction compared to the inclusive  
$^{12}$C$(K^-, K^+)$ reaction cross section indicates that the $\Xi^-$ emission probability 
peaks at 67\% in the region of $E_\Xi=100$ to 150 MeV. 
The remaining 33\% of $\Xi^-$ particles are involved in
$\Xi^-p$ inelastic scattering processes leading to
$\Xi^0n$ charge exchange and $\Lambda\Lambda$ conversion reactions. 
A classical approach using eikonal approximation shows that the total cross sections 
for $\Xi^{-}p$ inelastic scattering range between 42 mb and 23 mb in the $\Xi^-$ momentum 
range from 0.4 to 0.6 GeV$/c$. 
Furthermore, the differential cross section 
for $^{12}$C$(K^-, K^+\Lambda\Lambda)$ reaction indicates that 
the $\Xi^-p \to \Lambda\Lambda$ contributes to 57\%
in the energy range of $E_\Xi=0$ to 150 MeV.
The remaining contributions arises from two-step processes involving 
intermediate mesons such as $\pi^0$, $\eta$, $\omega$ and $\eta^\prime$, 
as determined by intranuclear cascade model calculations. 
Additionally, the total cross sections for the $\Xi^-p\to\Lambda\Lambda$ reaction 
are estimated in the same approach to range from 2.2 mb to 1.0 mb 
in the momentum range of 0.40 to 0.65 GeV$/c$. 
In particular, the momentum region between 0.5 and 0.6 GeV$/c$
corresponds to the typical momentum values of the $\Xi^-$ particles produced 
in the $^{12}$C$(K^-, K^+)$ reaction at 1.8 GeV$/c$, especially 
for $K^+$ scattering angles of less than $5^\circ$.
In this momentum interval, the cross section for the $\Xi^-p\to\Lambda\Lambda$ reaction is 
measured to be 1.0 mb.
This measurement result imposes a constraint on the upper bound of the decay width 
of the $\Xi^-$ particle in infinite nuclear matter, 
indicating that $\Gamma_\Xi< \sim 0.6$ MeV. 

\section*{Acknowledgment}

We thank the J-PARC staff for the excellent operation of the accelerator. We acknowledge
support from National Research Foundation (NRF) of Korea 
(Grant No. 2020R1A3B2079993, RS-2024-00436392, RS-2024-0035518); 
the Ministry of Education, Culture, Sports, Science and Technology (MEXT) of Japan
(Grant No. 24105002, H1805403, 18H03706, 21H01097, 23K20852, and 21H00130).
 

%\bibliographystyle{ptephy}
%\bibliography{sample}
%
% once the .bbl file has been generated then place the text in your article.

%This is added by T. Yoneya (editor-in-chief) on 2020/07/09.

\let\doi\relax

%without this code before the command "\begin{thebibliography}{}" , an error will be %flagged. When the bibliography is provided as separate .bib file, then this code %should be placed above the commands "\bibliographystyle{}" and "\bibliography{}" %inside the main TeX file. 

\end{document}